\documentclass[lettersize,journal]{IEEEtran}
\usepackage{amsmath,amsfonts}
\usepackage{algorithmic}
\usepackage{algorithm}
\usepackage{array}
\usepackage[caption=false,font=normalsize,labelfont=sf,textfont=sf]{subfig}
\usepackage{textcomp}
\usepackage{stfloats}
\usepackage{url}
\usepackage{verbatim}
\usepackage{graphicx}
\usepackage{cite}
\usepackage{multirow}
\usepackage{graphicx}
\usepackage{tabularx}
\usepackage{booktabs}
\usepackage{xcolor}
\usepackage{url}
\begin{document}

\title{Sea-Undistort: A Dataset for Through-Water Image Restoration in High Resolution Airborne Bathymetric Mapping}

\author{Maximilian~Kromer, Panagiotis~Agrafiotis, %
        and Beg{\"u}m~Demir,~\IEEEmembership{Senior Member, IEEE}
\thanks{This work is part of MagicBathy project funded by the European Union’s HORIZON Europe research and innovation programme under Marie Skłodowska-Curie Actions with Grant Agreement No. 101063294. Special thanks to Prof. D. Skarlatos (CUT) for research discussions and for providing the SfM-MVS data. \textit{(Corresponding author: Panagiotis Agrafiotis)}

The authors are with the Faculty of Electrical Engineering and Computer Science, Technische Universit{\"a}t Berlin, Germany. P. Agrafiotis and B. Demir are also with the Berlin Institute for the Foundations of Learning and
Data (BIFOLD), 10623 Berlin, Germany (e-mail: m.kromer@tu-berlin.de; agrafiotis@tu-berlin.de; demir@tu-berlin.de).}
}

\markboth{Journal of \LaTeX\ Class Files,~Vol.~14, No.~8, August~2021}%
{Shell \MakeLowercase{\textit{et al.}}: A Sample Article Using IEEEtran.cls for IEEE Journals}


\maketitle

\begin{abstract}
Accurate image-based bathymetric mapping in shallow waters remains challenging due to the complex optical distortions such as wave induced patterns, scattering and sunglint, introduced by the dynamic water surface, the water column properties, and solar illumination. In this work, we introduce Sea-Undistort, a comprehensive synthetic dataset of 1200 paired 512×512 through-water scenes rendered in Blender. Each pair comprises a distortion‐free and a distorted view, featuring realistic water effects such as sun glint, waves, and scattering over diverse seabeds. Accompanied by per‐image metadata such as camera parameters, sun position, and average depth, Sea-Undistort enables supervised training that is otherwise infeasible in real environments. We use Sea-Undistort to benchmark two state-of-the-art image restoration methods alongside an enhanced lightweight diffusion‐based framework with an early‐fusion sun-glint mask. When applied to real aerial data, the enhanced diffusion model delivers more complete Digital Surface Models (DSMs) of the seabed, especially in deeper areas, reduces bathymetric errors, suppresses glint and scattering, and crisply restores fine seabed details. Dataset, weights, and code are publicly available at https://www.magicbathy.eu/Sea-Undistort.html.
\end{abstract}


\section{Introduction}
\IEEEPARstart{A}{lthough} approximately 71\% of the Earth's surface is covered by water, with oceans accounting for around 96.5\% of that, only a limited portion of the seabed has been systematically mapped. This gap is critical, since detailed bathymetry underpins environmental monitoring, disaster response, heritage preservation, navigation safety, and offshore resource management. In shallow coastal regions, which are particularly susceptible to both natural and anthropogenic stressors, high-resolution bathymetry is especially important. Traditional mapping systems such as echo sounders often perform poorly in shallow waters due to wave interference, bottom clutter, and multipath errors. LiDAR systems, while useful, are expensive and generally lack the ability to provide semantic-rich data products \cite{agrafiotis2020}. To overcome these limitations, there has been growing interest in the use of aerial and satellite imagery for shallow-water bathymetric mapping \cite{agrafiotis2025deep}. Techniques such as Structure-from-Motion and Multi-view Stereo (SfM-MVS) with refraction correction, spectrally derived bathymetry (SDB), and hybrid approaches \cite{agrafiotis2025deep}  offer promising alternatives to traditional solutions. SfM reconstructs 3D structures using overlapping images captured from different viewpoints. In bathymetry, SfM paired with MVS generates shallow water 3D models from aerial or satellite imagery \cite{agrafiotis2020, agrafiotis2025deep}. Although well‐established on land, throught-water applications face significant optical distortions, especially refraction, which must be corrected for highly detailed and accurate bathymetric reconstructions. However, their performance is still strongly influenced by the quality of the input imagery and the inherent complexity of the underwater scene, limiting the maximum retrievable depth and DSM coverage \cite{agrafiotis2025deep}. SDB methods use satellite or aerial imagery and model radiance attenuation through the water column to estimate depth across shallow‐water regions, independent of seabed texture. Traditional approaches include empirical models and physics‐informed or machine learning–based regressions \cite{eugenio2021high}. Recently, deep learning models have shown a superior ability to capture complex light-water interactions \cite{zhou2023bathymetry, agrafiotis2025deep}, although their performance still depends heavily on input image quality.

In the literature, some approaches have been proposed for restoration of throughwater images, mainly to mask or correct sun glint \cite{giles2023combining, tivskus2023evaluation, qin2024advancing, pak2025sun}. However, these methods typically overlook the remaining dynamic optical distortions introduced by the aquatic environment, such as wave-induced surface deformation and light scattering within the water column. These phenomena significantly impact the radiometric and geometric quality of remotely sensed imagery, particularly in shallow water conditions. Another recent method \cite{scilla2025computer} operates on video frames and applies color transfer algorithms and image averaging to mitigate the optical distortions caused by both refraction and sun glint. A major limitation in the development of correction and enhancement methods lies in the absence of a comprehensive dataset that captures all of these water-related distortions in a controlled and paired manner. Acquiring such data in the real conditions is extremely difficult, if not impossible, due to the inability to capture perfectly aligned overwater scenes with and without distortions under the same environmental conditions. This lack of ground-truth-paired imagery has hindered the development and evaluation of learning-based methods that aim to correct or improve especially the high resolution through-water imagery.

Despite substantial progress in bathymetric retrieval-from SfM-MVS with refraction correction to deep learning-based SDB methods-challenges persist due to the complex optical properties of aquatic environments. To address this, we introduce Sea-Undistort, a novel synthetic dataset simulating high-resolution through-water scenes with and without distortions, enabling supervised training in the absence of real paired data. We conduct a new deep image restoration study on both synthetic and real imagery, comparing baseline and adapted models, including our diffusion-based enhanced model. Our results show that image restoration improves radiometric consistency, bathymetric coverage, and accuracy when integrated into SfM-MVS and learning-based SDB pipelines, respectively.

The remainder of this letter is structured as follows: Section \ref{sec:Sea-Undistort Dataset} introduces the Sea-Undistort dataset; Section \ref{sec:experiments} presents the experimental setup and results; Section \ref{sec:conclusion} concludes.

\section{The Sea-Undistort Dataset}
\label{sec:Sea-Undistort Dataset}
\begin{figure}
  \setlength{\tabcolsep}{1.5pt}
  \renewcommand{\arraystretch}{1}
  \footnotesize
\centering
  \begin{tabular}{cccc}
 \begin{minipage}[c]{0.235\columnwidth}
        \centering
        \includegraphics[width=\linewidth]{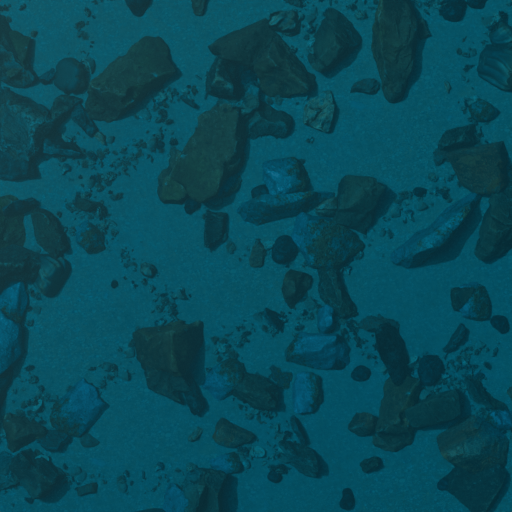}
    \end{minipage}& 
    
    \begin{minipage}[c]{0.235\columnwidth}
        \centering
        \includegraphics[width=\linewidth]{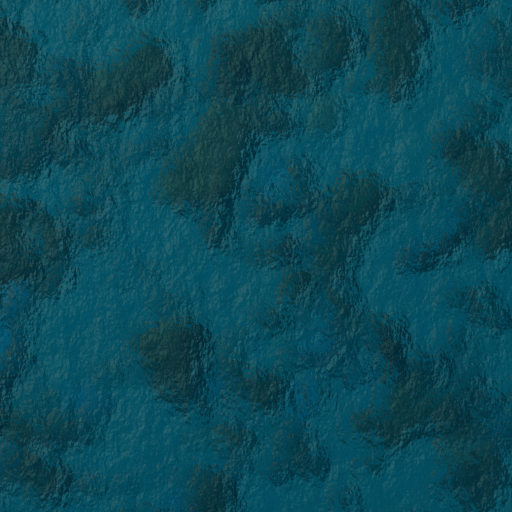}
    \end{minipage}& 
    
    \begin{minipage}[c]{0.235\columnwidth}
        \includegraphics[width=\linewidth]{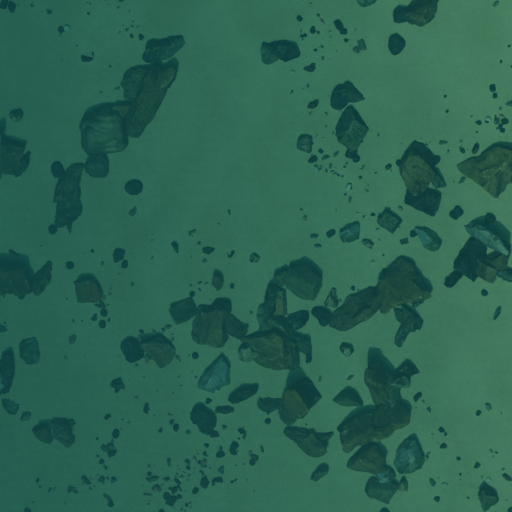}
    \end{minipage}&
    
            \begin{minipage}[c]{0.235\columnwidth}
        \includegraphics[width=\linewidth]
        {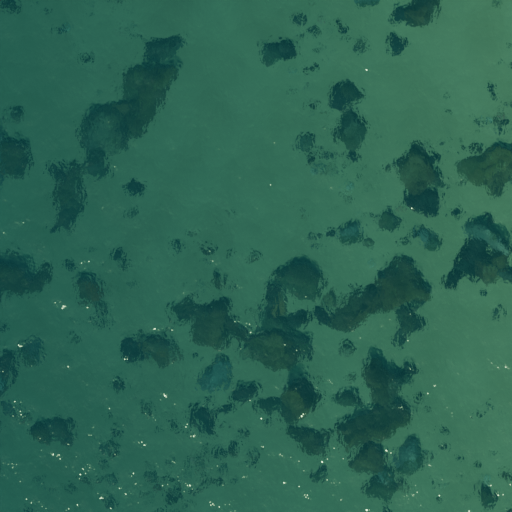}
    \end{minipage}\vspace{1pt}\\

          \begin{minipage}[c]{0.235\columnwidth}
          \centering
        \includegraphics[width=\linewidth]{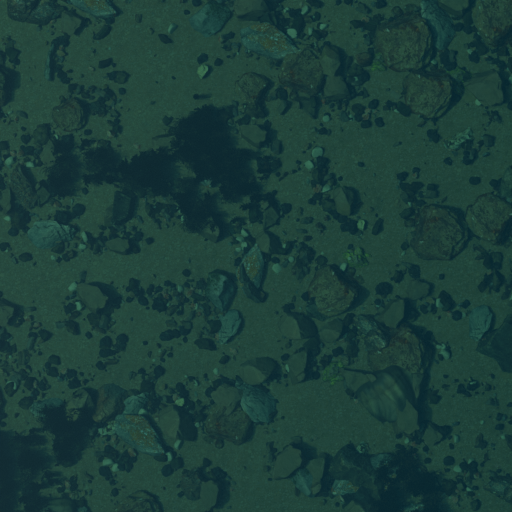}
        \textbf{(a)}
    \end{minipage}&
            
        \begin{minipage}[c]{0.235\columnwidth}
        \centering
        \includegraphics[width=\linewidth]{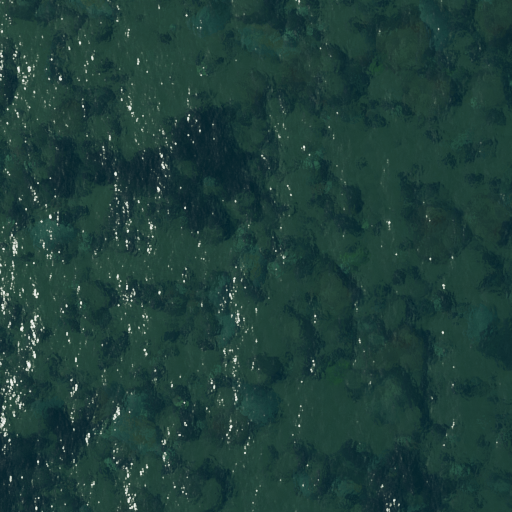}
        \textbf{(b)}    
    \end{minipage}&
    
        \begin{minipage}[c]{0.235\columnwidth}
        \centering
        \includegraphics[width=\linewidth]{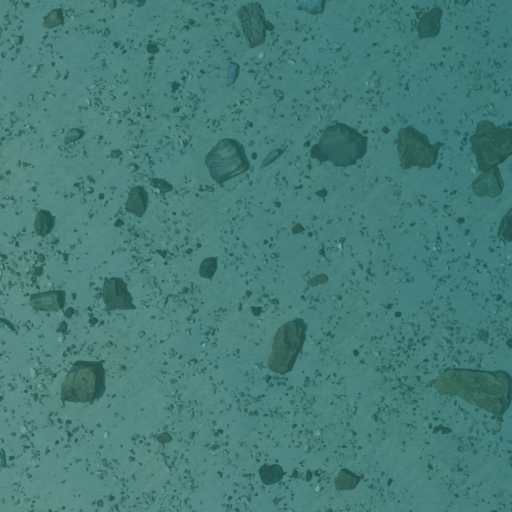}
        \textbf{(c)}
    \end{minipage}&
    
        \begin{minipage}[c]{0.235\columnwidth}
        \centering
        \includegraphics[width=\linewidth]{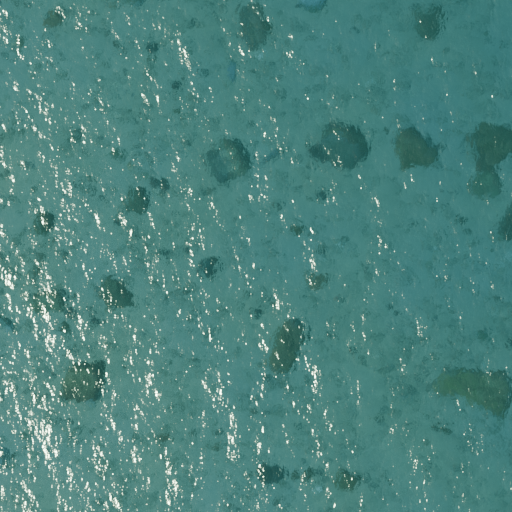}
        \textbf{(d)}
    \end{minipage}\vspace{1pt}\\

\end{tabular}

\caption{Example patches of the Sea-Undistort dataset without (a and c) and with (b and d) the dynamic optical distortions introduced by the water.}
\label{fig:fig1}
\end{figure}

To address the absence of real paired imagery with and without wave-, water-induced distortions, and sun glint, we introduce Sea-Undistort, a synthetic dataset created using the open-source 3D graphics platform Blender. The dataset comprises 1200 image pairs, each consisting of 512×512 pixel RGB renderings of shallow underwater scenes. Every synthetic pair includes a “non-distorted” image, representing minimal surface and water column distortions, and a corresponding “distorted” version that incorporates realistic optical phenomena such as sun glint, wave-induced deformations, and light scattering (Fig.\ref{fig:fig1}). These effects are procedurally generated to replicate the diverse challenges encountered in through-water imaging. The scenes are designed with randomized combinations of typical shallow-water seabed types, including rocky outcrops, sandy flats, gravel beds, and seagrass patches, capturing a wide range of textures, reflectance patterns, and radiometric conditions. Refraction is accurately modeled in both the distorted and non-distorted images to maintain geometric consistency with real underwater imaging physics.

In addition, camera settings are uniformly sampled within specific ranges to ensure diverse imaging conditions. Sensor characteristics include a physical width of 36 mm and effective pixel widths of 4000 or 5472 pixels. Focal lengths of 20 mm and 24 mm are simulated with only the central 512x512 pixels rendered. Camera altitude ranges from 30 m to 200 m, resulting in a ground sampling distance (GSD) between 0.014 m and 0.063 m. Average depths range from –0.5 m to –8 m, with a maximum tilt angle of 5°. Sun elevation angles between 25° and 70°, along with varying atmospheric parameters (e.g., air, dust), are used to simulate different illumination conditions. Generated images are accompanied by a ".json" file containing this metadata per image. Sea-Undistort is designed to support supervised training of deep learning models for through-water image enhancement and correction, enabling generalization to real-world conditions where undistorted ground truth is otherwise unobtainable.

\section{Experiments}
\label{sec:experiments}
We exploit Sea-Undistort to train state-of-the-art methods for synthetic and real image restoration applications.
\subsection{Image Restoration Models}
To evaluate the effectiveness of Sea-Undistort, we performed qualitative and quantitative comparisons against two state-of-the-art baselines: the original ResShift model \cite{yue2024efficient} (a lightweight diffusion model), and NDR-Restore \cite{yao2024neural} (a neural degradation representation learning model). We further include an extended version of ResShift, referred to as ResShift+EF, which integrates early fusion (EF) between RGB imagery and a continuous sun-glint mask. This mask is derived from the V and S channels in HSV space: pixels above an upper threshold are set to 1 (strong glint), those below a lower threshold to 0 (no glint), and intermediate values are linearly interpolated. By incorporating both brightness and saturation cues, this one-channel map captures white, de-saturated areas indicative of intense glint. For EF, the sun-glint mask is concatenated as an additional channel to the RGB input, forming a four-channel input tensor that is then fed into the original ResShift model. This guides the diffusion process to better focus corrections on glint-affected regions, improving restoration of underwater features without introducing artifacts.

\subsection{Implementation Details}
We implemented the models in PyTorch using one NVIDIA A100 80GB GPU. The samples of Sea-Undistort were randomly divided into a training set of 802 samples (80\%), a validation set of 100 samples (10\%), and a test set of 100 samples (10\%). Some images from the initial 1200 were discarded due to overly intense optical effects that could bias the training. NDR-Restore used 128×128 images, L1 loss, the Adam optimizer, a learning rate of \(1\times10^{-4}\), and a MultiStepLR scheduler over 4630 epochs (~80 hours, ~37M parameters). In contrast, ResShift and ResShift+EF were trained with 256×256 images using a composite loss (latent-MSE, pixel-LPIPS, pixel-MSE; weights 1.0, 4.0, 1.0), AdamW without weight decay, a learning rate of \(5\times10^{-5}\), and a cosine scheduler with 500 warm-up steps. All models were trained for up to 200K iterations ($\simeq$72 hours each) on the same hardware setup, with 118.7 million parameters per model.). For inference on 100 images (512×512), NDR-Restore required 54 seconds, ResShift+EF 63 seconds, and ResShift 67 seconds.

\subsection{Results on Sea-Undistort (Synthetic) Data}
We first evaluate the trained models on the Sea-Undistort test set to assess performance under controlled conditions.

\subsubsection{Qualitative Comparison} 
\begin{figure}[h!]
  \setlength{\tabcolsep}{1.5pt}
  \renewcommand{\arraystretch}{1}
  \footnotesize
\centering
  \begin{tabular}{ccccc}
 \begin{minipage}[c]{0.235\columnwidth}
        \centering
        \includegraphics[width=\linewidth]{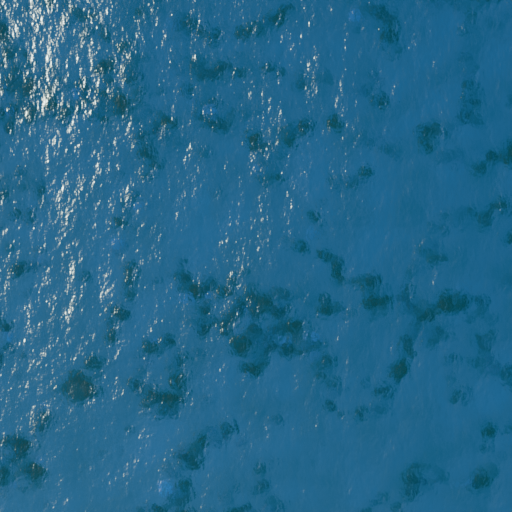}
    \end{minipage}& 
    
    \begin{minipage}[c]{0.235\columnwidth}
        \centering
        \includegraphics[width=\linewidth]{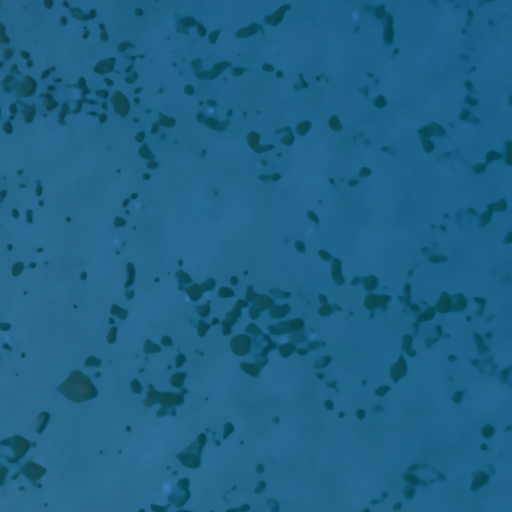}
    \end{minipage}& 
    
    \begin{minipage}[c]{0.235\columnwidth}
        \includegraphics[width=\linewidth]{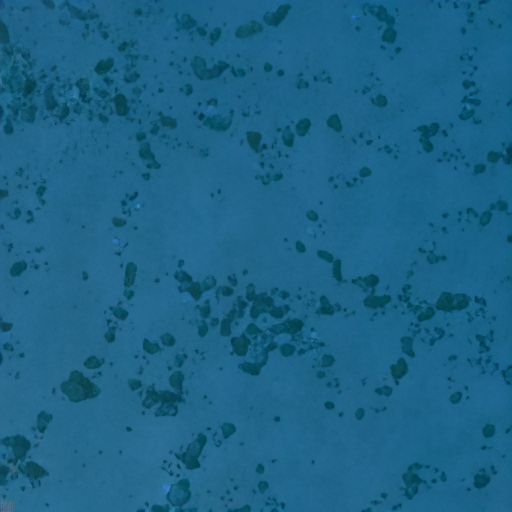}
    \end{minipage}&

          \begin{minipage}[c]{0.235\columnwidth}
        \includegraphics[width=\linewidth]{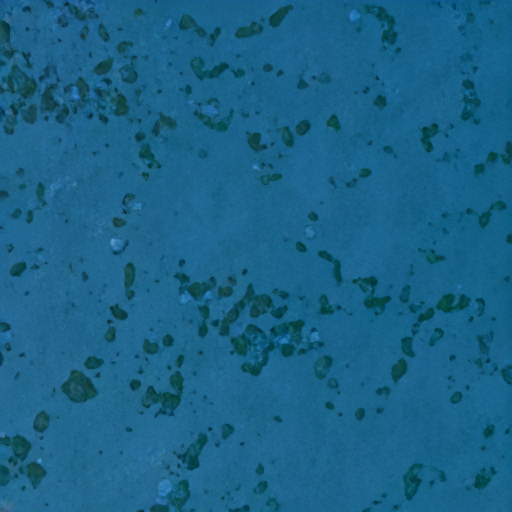}
    \end{minipage}&\vspace{1pt}\\

     \begin{minipage}[c]{0.235\columnwidth}
        \centering
        \includegraphics[width=\linewidth]{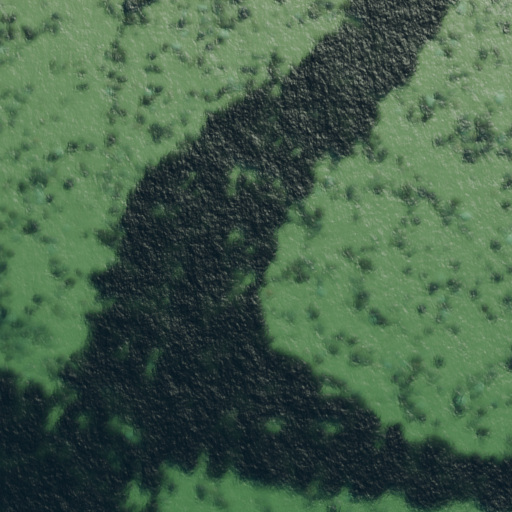}
        \textbf{(a)}
    \end{minipage}& 
    
    \begin{minipage}[c]{0.235\columnwidth}
        \centering
        \includegraphics[width=\linewidth]{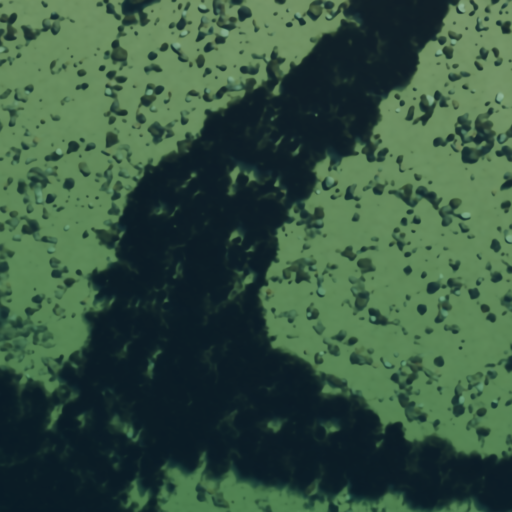}
        \textbf{(b)}
    \end{minipage}& 
    
    \begin{minipage}[c]{0.235\columnwidth}
    \centering
        \includegraphics[width=\linewidth]{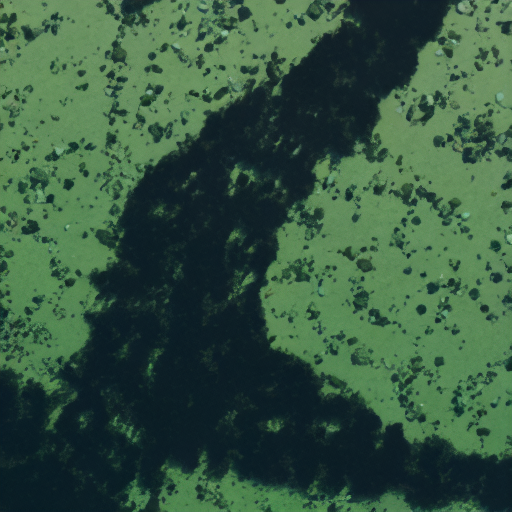}
        \textbf{(c)}
    \end{minipage}&

          \begin{minipage}[c]{0.235\columnwidth}
          \centering
        \includegraphics[width=\linewidth]{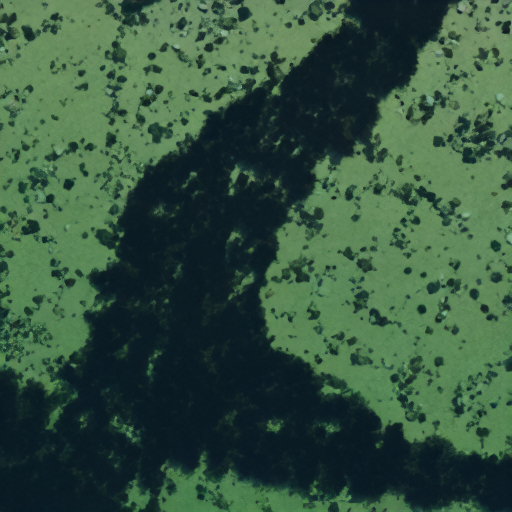}
        \textbf{(d)}
    \end{minipage}&
\end{tabular}
\caption{Example Sea-Undistort: (a) original patches; restorations using (b) NDR-Restore, (c) ResShift, and (d) ResShift+EF.}
\label{fig:fig2}
\end{figure}

Fig. \ref{fig:fig2} shows the original patches and restorations using NDR-Restore, ResShift, and ResShift+EF. There, one can observe that the original distorted patches exhibit pronounced sun-glint and scattering that obscure fine seabed textures and warp local contrast. NDR-Restore effectively suppresses large glint regions and reduces blurriness at the cost of over-smoothing small-scale details, resulting in a more homogeneous appearance. Vanilla ResShift preserves edge crispness and subtle textures more faithfully than NDR-Restore, though residual glint and mild blurriness remain, particularly in darker areas. By incorporating an early-fusion sunglint mask, ResShift+EF more completely mutes strong glint spots and reduces scattering without sacrificing the sharpness of undistorted regions, yielding the best visual balance between artifact removal and detail preservation.

\subsubsection{Quantitative Comparison} Since the Sea-Undistort provides paired distorted and non-distorted images, we also report reference-based image quality metrics such as SSIM (Structural Similarity Index) \cite{wang2004image}, PSNR (Peak Signal-to-Noise Ratio), LPIPS \cite{zhang2018lpips}, and two state-of-the-art no-reference methods specifically designed to assess image realism: CLIPIQA \cite{wang2023clipiqa} and MUSIQ (koniq-10k) \cite{ke2021musiq}, as described and used in \cite{yue2024efficient}. Table \ref{ssim-psnr} quantifies observations on Fig. \ref{fig:fig2}: NDR-Restore achieves the highest SSIM and PSNR, reflecting its strong pixel-level fidelity to the ground truth. However, its relatively high LPIPS score and lower perceptual ratings from CLIPIQA and MUSIQ reveal a loss of fine texture and the introduction of smoothing artifacts. Vanilla ResShift offers a clear perceptual improvement over NDR-Restore, with a lower LPIPS and higher CLIPIQA and MUSIQ, though its SSIM and PSNR remain below NDR-Restore’s. ResShift+EF enhances the perceptual fidelity of vanilla ResShift, achieving the lowest LPIPS, highest CLIPIQA, and MUSIQ scores while maintaining SSIM and PSNR values close to those of vanilla ResShift. This demonstrates that the early-fusion mask effectively guides restoration to problematic regions, refining texture realism without overly compromising pixel-wise accuracy, yielding an ideal trade-off for downstream SfM–MVS and learning-based bathymetric applications.

\begin{table}[t]
  \renewcommand{\arraystretch}{1}
  \footnotesize
  \caption{Image Quality Metrics for Sea-Undistort Testing Data Across Restoration Methods. Entities in Bold Indicate the Best Score. (↓: Lower is Better, ↑: Higher is Better)}
  \centering
  \label{ssim-psnr}
  \begin{tabular}{lrrrr}
    \toprule
    Metric & Original & NDR-Restore & ResShift & ResShift+EF \\
    \midrule
    SSIM ↑      & 0.662    & \textbf{0.854}     & 0.771         & 0.756    \\
    PSNR ↑      & 28.920   & \textbf{33.950}    & 31.120        & 30.570     \\
    LPIPS ↓     & 0.377    & 0.266              & 0.143         & \textbf{0.112}    \\
    CLIPIQA ↑   & 0.316    & 0.357              & 0.405         & \textbf{0.414}    \\
    MUSIQ ↑     & 40.493   & 36.425             & 40.330        & \textbf{41.235}  \\
    \bottomrule
  \end{tabular}
\end{table}

\subsection{Results on Real Data}
To evaluate the models trained using Sea-Undistort on real data we conducted testing on two distinct types of datasets corresponding to the two primary image-based bathymetric retrieval approaches: SfM-MVS and learning-based SDB. This was necessary since ground truth imagery is not available for evaluation in real applications. For the learning-based SDB evaluation, we used the aerial modality of the MagicBathyNet dataset \cite{magicbathynet} which is specifically curated to train and test deep learning models for shallow water bathymetry. The aerial subset includes RGB orthoimagery (0.25 m GSD) paired with corresponding DSMs of the seabed derived from high-accuracy LiDAR and SONAR systems, allowing quantitative performance evaluation. To evaluate the SfM-MVS-based approach, we used the photogrammetric blocks of refraction-corrected \cite{agrafiotis2021} low-altitude aerial imagery used to build MagicBathyNet \cite{magicbathynet}. These blocks cover shallow‐water areas with varied seabed compositions (sand, rock, macroalgae, seagrass) and depths (0 m to –20 m). Imagery is captured from an average flying height of 209 m, resulting in a 0.063 m average GSD. They also exhibit differing water‐column and surface properties (turbidity, clarity, waves, sun-glint etc.) and include detailed reference bathymetric data and acquisition parameters. Originally employed in \cite{agrafiotis2020}, they provide a reliable benchmark. All necessary details regarding the dataset, including acquisition and processing, are provided in \cite{magicbathynet}.

\subsubsection{Visual Comparison}
\begin{figure}[h!]
  \setlength{\tabcolsep}{1.5pt}
  \renewcommand{\arraystretch}{1}
  \footnotesize
\centering
  \begin{tabular}{cccc}
    \begin{minipage}[c]{0.235\columnwidth}
          \centering
        \includegraphics[width=\linewidth]{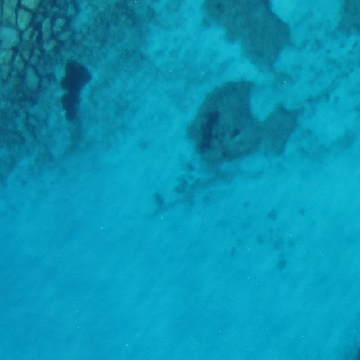}

    \end{minipage}&
            
        \begin{minipage}[c]{0.235\columnwidth}
        \centering
        \includegraphics[width=\linewidth]{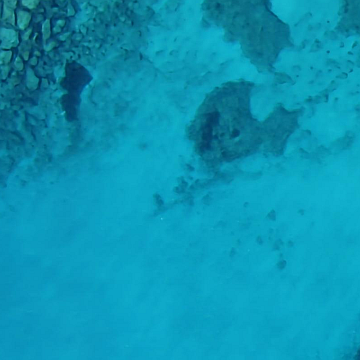}

    \end{minipage}&
    
        \begin{minipage}[c]{0.235\columnwidth}
        \includegraphics[width=\linewidth]{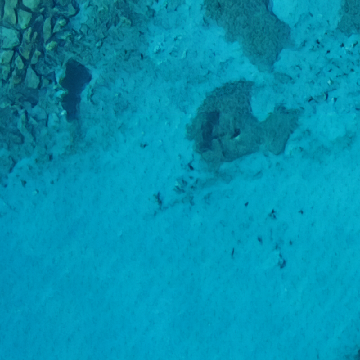}
        \centering

    \end{minipage}&
    
        \begin{minipage}[c]{0.235\columnwidth}
        \centering
        \includegraphics[width=\linewidth]{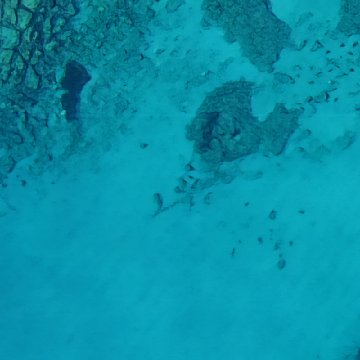}
    
    \end{minipage}\vspace{1pt}\\

    \begin{minipage}[c]{0.235\columnwidth}
          \centering
        \includegraphics[width=\linewidth]{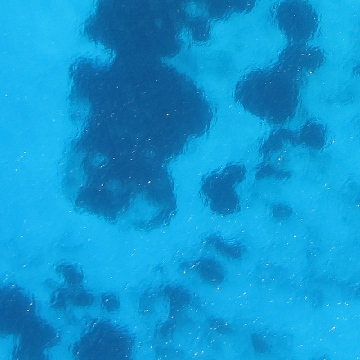}
        \textbf{(a)}
    \end{minipage}&
            
        \begin{minipage}[c]{0.235\columnwidth}
        \centering
        \includegraphics[width=\linewidth]{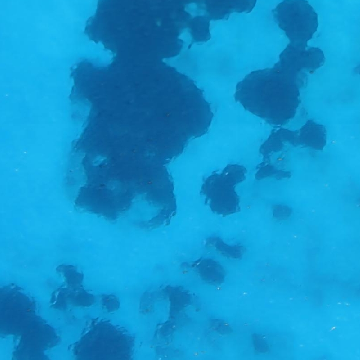}
        \textbf{(b)}
    \end{minipage}&
    
        \begin{minipage}[c]{0.235\columnwidth}
        \includegraphics[width=\linewidth]{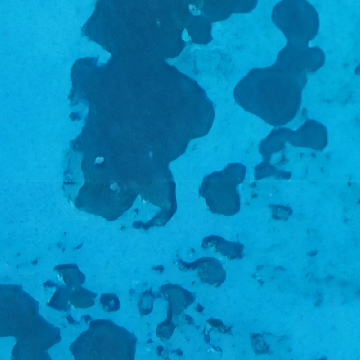}
        \centering
        \textbf{(c)}
    \end{minipage}&
    
        \begin{minipage}[c]{0.235\columnwidth}
        \centering
        \includegraphics[width=\linewidth]{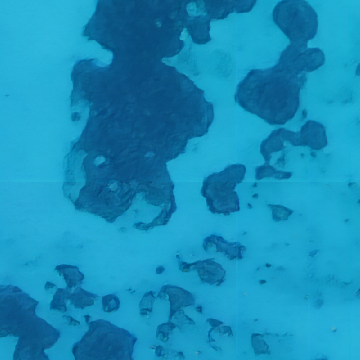}
        \textbf{(d)}
    \end{minipage}
    
\end{tabular}

\caption{Example of real imagery: (a) original patches; restorations using (b) NDR-Restore, (c) ResShift, and (d) ResShift+EF.}

\label{fig:fig3}
\end{figure}

Fig. \ref{fig:fig3} presents a qualitative comparison between the raw aerial images and the outputs of NDR-Restore, ResShift, and our ResShift+EF variant, focusing on the removal of surface and subsurface distortions, such as sun glint, wave-induced texture distortion, and scattering across various shallow-water conditions. The unprocessed aerial patches exhibit muted seabed textures due to water absorption, scattering, and sun glint. Although all restoration methods noticeably reduce blurriness and improve contrast, NDR-Restore smooths the fine rock fissures and gravel patterns, and vanilla ResShift better preserves edge sharpness and natural color balance (although residual blurry remains in deeper regions). Our ResShift+ER delivers the best overall result. By effectively muting surface glint and subsurface scattering without over-saturation, it yields the crispest recovery of small-scale seabed features (e.g. cracks, seagrass outlines) and maintains natural color fidelity. This balance of clarity, texture preservation, and contrast consistency underscores the benefit of incorporating the early contextual fusion step. This comparison highlights improvements in visual clarity, contrast consistency, and restoration of seabed textures critical for downstream bathymetric processing.

\subsubsection{SfM-MVS-based Bathymetry Comparison}  For each photogrammetric block, we processed the original and restored images and generated DSMs at 0.25 m GSD in order to quantify how image restoration impacts reconstruction completeness, the maximum depth reached, and bathymetric accuracy. DSMs provide a uniform representation that is directly comparable and is widely used in SfM‐MVS evaluations without relying on irregular point‐cloud densities and excluding any photogrammetric noise. Specifically, we compare the number of DSM pixels generated by each restoration method across predefined 2 m depth bins (Table \ref{tab:point_counts}). No interpolation is applied, and pixels or areas without a reconstructed 3D point remain empty. This ensures unbiased results that reflect how well each enhancement method preserves features critical for through‐water seabed 3D reconstruction and DSM generation.

\begin{table}[h!]
\footnotesize
\caption{DSM Pixel Counts per Depth Bin for Different Restoration Methods. Entities in Bold Indicate the Best Score.}
\label{tab:point_counts}
\centering
\begin{tabular}{lrrrr}
\toprule
Depth Bin (m) & Original & NDR-Restore & ResShift & ResShift+EF \\
\midrule
0 to –2      & 1197373 &          1282498  & 1044499  & \textbf{ 1297283} \\
–2 to –4     & 2048178 & \textbf{ 2140408} & 1801064  &          2066663  \\
–4 to –6     & 2615469 & \textbf{ 2734804} & 2165789  &          2574398  \\
–6 to –8     &  628867 & \textbf{  684114} &  404654  &           590727  \\
–8 to –10    &  307962 & \textbf{  360168} &  149352  &           273316  \\
–10 to –12   &  229988 & \textbf{  313068} &  105945  &           225009  \\
–12 to –14   &  157263 & \textbf{  270698} &   89307  &           185263  \\
–14 to –16   &   37515 & \textbf{   91783} &    21574 &  69812\\
–16 to –18   &    2756 &              8664 &     2508 & \textbf{    9161} \\
–18 to –20   &       2 &                14 &       31 & \textbf{      46} \\
\bottomrule
\end{tabular}
\end{table}

Excluding noise, a higher DSM pixel count per bin indicates improved image quality, as it suggests that more features were reliably matched and triangulated, yielding denser reconstructions and expanding coverage - provided that accuracy is preserved. Several trends emerge: \textit{Shallow depths (0 to –6 m):} ResShift+EF yields the most DSM pixels in 0 to –2 m, indicating that it reveals additional shallow features in the wave-braking affected zone. In –2 to –6 m, NDR-Restore leads, with ResShift+EF a close second, showing both boost feature detectability in moderately shallow water. \textit{Mid depths (–6 to –14 m):} NDR-Restore consistently produces the highest counts, recovering more mid-range seabed details. The original imagery trails slightly, while ResShift+EF and ResShift offer fewer pixels but likely higher accuracy, reflecting a trade-off between density and noise suppression. \textit{Greater depths (–14 to –20 m):} From –14 to –16 m, NDR-Restore again tops counts, with ResShift variants performing similarly. Below –16 m, ResShift+EF reports higher counts (9161 at –16 to –18 m; 46 at –18 to –20 m), reflecting its superior restoration capabilities under intense through-water distortions. Other methods register less DSM pixels in the deepest bin, consistent with less restored imagery at those depths.

Table \ref{sfm depths} shows the DSM accuracy results compared to the reference bathymetry. It is evident that despite substantially extending the bathymetric coverage into deeper seabed areas, ResShift+EF maintains comparable bathymetric accuracy to the DSM derived using the non-restored imagery while ResShift slightly improves it. NDR-Restore results in a moderate decline.  Depth accuracy is assessed using Root Mean Square Error (RMSE), Mean Squared Error (MSE), and Standard Deviation (STD). 

\begin{table}[h!]
  \renewcommand{\arraystretch}{1}
  \footnotesize
  \caption{DSM Accuracy Results Compared to Reference Bathymetry. Entities in Bold Indicate the Best Score.}
  \centering
  \label{sfm depths}
  \begin{tabular}{lccc}
    \toprule
    Method & RMSE (m) & MAE (m) & STD (m) \\
    \midrule
    Original & 0.934 & 0.424 & 0.933  \\
    NDR-Restore & 1.099 & 0.474 & 1.098 \\
    ResShift & \textbf{0.827} & \textbf{0.388} & \textbf{0.826}    \\
    ResShift+EF& 0.995 & 0.434 & 0.995  \\
    \bottomrule
  \end{tabular}
\end{table}

This combination of extended coverage, indicating successful image matching in deeper areas, and maintained DSM accuracy indicates that the tested restoration models did not introduce significant hallucinations or spurious depth artifacts.

\subsubsection{Learning-based Bathymetry Comparison} We applied our image restoration models to input orthoimagery prior to training and compared the depth prediction results with the ground truth \cite{magicbathynet} to assess improvements in bathymetric accuracy (Table \ref{learning-based depths}). For training and inference using the restored imagery, we considered UNet-bathy \cite{magicbathynet}. The initial learning rate was 5x$10^{-6}$ for a 10 epoch training period. The learning rate decreased by a factor of 10 after 9 epochs. For this analysis, we focus exclusively on the Agia Napa area, which features greater depth variability and more pronounced water-induced distortions, making it well-suited to demonstrate the value of the proposed Sea-Undistort dataset in training and evaluating restoration models.

\begin{table}[h!]
  \renewcommand{\arraystretch}{1}
  \footnotesize
  \caption{Accuracy Obtained by Learning-based Bathymetry Method. Entities in Bold Indicate the Best Score.}
  \centering
  \label{learning-based depths}
  \begin{tabular}{lccc}
    \toprule
    Method & RMSE (m) & MAE (m) & STD (m) \\
    \midrule
    Original & 0.616 & 0.420 & 0.598  \\
    NDR-Restore & 0.576 & 0.392 & 0.582  \\
    ResShift & 0.596 & 0.488 & 0.582   \\
    ResShift+EF& \textbf{0.556} & \textbf{0.336} & \textbf{0.515}  \\

    \bottomrule
  \end{tabular}
\end{table}

\subsubsection{Quantitative Comparison} 
Since ground-truth undistorted references are unavailable, we evaluate perceptual improvements using a comprehensive set of no-reference image quality metrics. These include: BRISQUE \cite{brisque}, NIQE \cite{niqe}, 
and PIQE \cite{piqe}, which are traditional perceptual quality estimators, Sharpness and Entropy which are indicators of texture preservation and information richness, MUSIQ (ava) and MUSIQ (koniq-10k) \cite{ke2021musiq}, which are multi-scale quality scores based on learned human perceptual judgments, and CONTRIQUE \cite{Madhusudana2022contrique}, which is a content-aware deep learning-based quality assessment model. These metrics provide an objective basis for comparing restoration performance across methods under uncontrolled, real-world aquatic conditions. Table \ref{tab:iqa_methods_final} shows that original images score best on NIQE, indicating the lowest predicted distortion before enhancement.

\begin{table}[ht]
\centering
\footnotesize
\setlength{\tabcolsep}{4pt}
\caption{No‐Reference Image Quality Metrics. Entities in Bold Indicate the Best Score. (↓: Lower is Better, ↑: Higher is Better)}
\label{tab:iqa_methods_final}
\begin{tabular}{lrrrr}
\toprule
Metric       & Original & NDR-Restore & ResShift & ResShift+EF \\
\midrule
BRISQUE ↓     & 34.350            & 39.190              & 30.940            & \textbf{29.490}          \\
NIQE ↓        & \textbf{3.080}    & 3.890               & 5.760             & 5.620                    \\
PIQE ↓        & 21.500            & 43.180              & \textbf{6.870}    & 13.060                   \\
Sharpness ↑   & 86.820            & 58.390              & 219.510           & \textbf{234.820}         \\
Entropy ↑     & 6.590             & 6.590               & \textbf{6.730}    & 6.600                    \\
MUSIQ (ava) ↑ & 4.190             & 4.201               & \textbf{4.291}    & 4.229                    \\
MUSIQ (koniq-10k) ↑ & 50.885      & 50.989              & 51.858            & \textbf{52.160}          \\
CONTRIQUE ↑   & 51.874            & 53.088              & \textbf{56.833}   & 54.820                   \\
\bottomrule
\end{tabular}
\end{table}

ResShift+EF achieves the lowest BRISQUE and highest Sharpness, demonstrating the strongest artifact removal and fine‑detail enhancement. While it reduces PIQE, vanilla ResShift has the lowest PIQE overall. ResShift further excels in Entropy, MUSIQ (ava), and CONTRIQUE, highlighting its texture richness and aesthetic quality.
Moreover, ResShift+EF outperforms ResShift on MUSIQ (koniq-10k), confirming its enhanced technical image quality performance. NDR-Restore, by contrast, offers a lower NIQE than either ResShift variant, but lags in sharpness and distortion metrics. Taken together, these results indicate that ResShift delivers the best texture‑and‑learned‑perceptual gains, ResShift+EF provides the strongest overall artifact suppression and detail refinement, and NDR‑Restore preserves naturalness more conservatively.

\section{Discussion and Conclusion}
\label{sec:conclusion}
In this work, we presented Sea-Undistort, a novel synthetic dataset tailored for supervised restoration of high-resolution through-water imagery under controlled yet realistic aquatic conditions. With 1200 paired distorted and non-distorted airborne through-water RGB scenes, Sea-Undistort enables systematic benchmarking and training of deep learning models where real paired data is unobtainable. This dataset fills a critical gap in the field by procedurally simulating complex surface and subsurface distortions, including sunglint, waves, turbidity, and scattering, while providing detailed metadata for each image. To demonstrate the utility of Sea-Undistort, we evaluated several image restoration models. Models trained on Sea-Undistort showed strong generalization to real airborne through-water imagery, improving seabed visibility and enhancing the quality of bathymetric products in both bathymetric SfM–MVS and learning-based SDB. As a future work, we plan to exploit metadata (e.g., sensor, environmental, and geometric information) in the learning process to enhance restoration accuracy and bathymetry retrieval further.

\bibliographystyle{IEEEtran}
\bibliography{IEEEabrv,refs}

\vfill

\end{document}